\documentstyle
{article}
\begin{document}
\newcommand{\bb}{\begin{eqnarray}}
\newcommand{\ee}{\end{eqnarray}}
\newcommand{\tn}{\tt \normalsize}
\newcommand{\n}{\newline}
\baselineskip=8.00truemm     \rm  \Large
\begin {center}\bf Physical States in Topological Phase\\
of N=2 Su\-per\-sym\-met\-ric WZNW Models \rm \end{center}
\large \rm
\vspace{10.00mm}
\begin {center}   S.Parkhomenko  \rm \end{center}
\vspace{5.00mm}
\begin {center}   L. D. Landau Institute of Theoretical Physics
of the Russian Academy of Sciences  \end{center}
\vspace{5.00mm}
\begin {center}   January 1993  \end{center}
\vspace{10.00mm}
\large \rm

   \bf Abstract. \rm
N=2 su\-per\-sym\-met\-ric WZNW mod\-els\\
as\-so\-ci\-ated
with finite-dimensional Manin
triples $(p,p_{+},p_{-})$, where $p_{\pm}$ are
Borel sub\-al\-geb\-ras of
any simple Lie algebra is considered.
Physical states in topological phase
of these models are computed. Them
ring stucture and N=2 WZNW
representatives are constracted.

 \vspace{15.00mm}

\bf 1. Introduction \rm   \large
The extended superconformal field theories
in two dimensions have recently at\-trac\-ted
attension for two reasons.
First, in connection with compactification
of strings with space-time supersymmetry [1-4].
Second, as was noted in [5]
the so-called twisted N=2 models introduced in [6-7]
describe topological matter
wich can interact with topological gravity.
{}From the point of view of classification of
conformal theiries, N=2 superconformal theories
are in asense the simplest type to classify,
and a subset of them, supersymmetric Landau-
Ginzburg theories, is related to catastrophe
theory [8]. From the point of view of topological
characterization of the theory, they have a ring
of operators (chiral primary fields) wich
is believed to basicaly characterize them.\n
In [9-11] WZNW models were studed
,wich allow extended supersymmetry,
and conditions were formulated that the Lie
group must sutisfy so that its WZNW model
would have extended supersymmetry.
In particular, in [11] a correspondence was established
between N=2,4 super
conformal Kac-Moody algebras
and finite-dimensional Manin triples.\n
A generators of Super-Virasoro algebras arising
in N=2,4 WZNW models are given
by appropriate supersymmetryc
extension of Sugavara construction.
In [12-16]
were studed a representations and modular properties
of characters of these algebras.\
In the present paper we consider the "topological phase"
of N=2 WZNW model wich
are assotiated with Manin triples
of the form $(p,p_{+},p_{-})$, where $p_{+}$,
$p_{-}$ are Borel subalgebras of any simple Lie algebra and
$p=p_{+}\oplus p_{-}$
as a vector space. Here the "topological phase" means
the topological theory wich
is obtained from N=2 WZNW model with help of twisting
procedure [6-7]. We compute
a physical states in topological phase.
The ring structure of these states will be
investigated in the next paper.\n
The paper is arranged as follows.
In section 2 we briefly describe Sugavara construction
following [10]. In section 3 we formulate the problem of describing
a physical states in topologycal phase as a current
algebra cohomology problem.
Section 4 is devoted to the solution of this problem.
In section 5 we give in an
explicit form a representatives of physical
states and show that they are a
highest weightes of N=2 Virasoro algebra.
In section 6 we compute the ring
structure of physical states. \n

\bf  2. N=2 WZNW models and finite-dimensional
Manin triples \rm   \large\n
Folowing [11] we give the N=2 modification of Sugavara construction.
Let $p$ be finite-dimensional Lie algebra, endowed with invariant nondegenerate
scalar product $<,>$, and $p_{+}$, $p_{-}$ be isotropic
with respect to this scalar
product Lie subalgebras of $p$, such that
$p=p_{+}\bigoplus p_{-}$ as vector
space. Then the triple $(p,p_{+},p_{-})$ is called Manin triple.Let's fixed any
finite-dimensional Manin triple and ortonormal basis $E^{a}$, $E_{a}$, $a=0,...
,\frac{dimp}{2}-1$ in $p$ such that $f^{ab}_{c}$, $f_{ab}^{c}$
are the structure constants
of subalgebras $p_{+}$, $p_{-}$. Let $J^{a}(z)$, $J_{a}(z)$ be a generators
of the
affine Kac-Moody algebra $\hat{Lp}$, wich are correspondes to the fixed
basis $E^{a}$,
$E_{a}$ and $J^{a}(z)$ carrents generate
$Lp_{+}$ subalgebra, $J_{a}(z)$
currents
generate $Lp_{-}$ subalgebra. Let $\psi^{a}(z)$,$\psi_{a}(z)$, $a=1,...
,\frac{dimp}{2}-1$
be a free fermion carrents wich has a singular OPE with respect to the
scalar product
$\langle,\rangle$. The two spin $\frac{3}{2}$ currents:
\bb G^{+}=\alpha(J^{a}\psi_{a}+
    \beta^{+}f^{ab}_{c}\psi_{a}\psi_{b}\psi^{c})  \ee
\bb G_{-}=\alpha(J_{a}\psi^{a}+\beta^{-}
    f_{ab}^{c}\psi^{a}\psi^{b}\psi_{c}) \ee
,where $\alpha$, $\beta^{+}$, $\beta^{-}$ are normalization constants,
generate N=2
Virasoro superalgebra.\n
The example of the Manin triple wich
will be important for us in the following
is based on any simple Lie algebra
$g$ with the scalar product $(,)$ and its Cartan
decomposition $g=n_{-}\oplus h\oplus n_{+}$, $b_{+}=
h\oplus n_{+}$, $b_{-}=h\oplus n_{-}$.
Consider the Lie algebra
\bb   p=g\oplus \tilde{h}      \ee
,where $\tilde{h}$ is the copy of the
Cartan subalgebra $h$ and the Lie algebra
structure on $p$ is defined by
\bb    [g,\tilde{h}]=0     \ee
On $p$ we define the invariant scalar product
\bb  <(X_{1},H_{1}),(X_{2},H_{2})>=(X_{1},X_{2})-(H_{1},H_{2})   \ee
If we set
\bb p_{+}=\{(X,H)\in p|X\in b_{+},H=X_{h}\}  \ee
\bb p_{-}=\{(X,H)\in p|X\in b_{-},H=-X_{h}\}  \ee
,where $X_{h}$ is the projection of $X$ on the Cartan subalgebra $h$,
then we have $p=p_{+}\oplus p_{-}$ and
$p_{+}$, $p_{-}$ are isotropic subalgebras of $p$,
wich are isomorphic to a Borel subalgebras $b_{+}$, $b_{-}$.
Remark that there is the direct generalization of this construction,
wich is based on any pair of parabolic sabalgebras $p_{\pm}$ such that
$b_{\pm}\subset p_{\pm}$.
The explicit N=2 Sugavara construction we give in the simplest case
\bb    g=sl(2,C)   \ee
For this case let $J^{a},J_{a},\psi^{a},\psi_{a}$, $a=0,1$
be the bosonic and fermionic carrents
with the OPE
\bb   J^{0}(z)J^{1}(0)=z^{-1}J^{1}(0)+o(z) \nonumber \\
      J_{0}(z)J_{1}(0)=z^{-1}J_{1}(0)+o(z) \nonumber \\
      J^{0}(z)J^{0}(0)=-z^{-2}+o(z)         \nonumber \\
      J_{0}(z)J_{0}(0)=-z^{-2}+o(z)         \nonumber \\
      J^{0}(z)J_{1}(0)=-z^{-1}J_{1}(0)+o(z) \nonumber \\
      J^{1}(z)J_{0}(0)=z^{-1}J^{1}(0)+o(z)  \nonumber \\
      J^{0}(z)J_{0}(0)=-z^{-2}(k+1)+o(z)    \nonumber \\
      J^{1}(z)J_{1}(0)=-z^{-2}k+z^{-1}(J_{0}-J^{0})(0)+o(z) \nonumber \\
      \psi^{a}(z)\psi_{b}(0)=z^{-1}\imath\delta^{a}_{b}+o(z)  \ee
If we introduce currents $H=J_{0}-J^{0}$, $P=J_{0}+J^{0}$ instead of
$J_{0}$,
$J^{0}$, then the OPE (9) is nothing less than OPE
of $\hat{Lsl(2)}$ current algebra
with level $k$ and OPE of $\hat{L\tilde{h}}$
current algebra with level $k+2$.
Generators of N=2 Virasoro superalgebra with
central charge $c=4\frac{k+1}{k+2}$ is given by
\bb T=-\frac{\imath}{2}(\partial\psi_{a}\psi^{a}-\psi_{a}\partial\psi^{a})-
    \frac{1}{2(k+2)}(J_{a}J^{a}+J^{a}J_{a})     \nonumber \\
    G^{-}=-\sqrt{\frac{2}{k+2}}(\psi^{a}J_{a}+
    \imath\psi^{0}\psi^{1}\psi_{1})           \nonumber \\
    G^{+}=-\sqrt{\frac{2}{k+2}}(\psi_{a}J^{a}+
    \imath\psi_{0}\psi_{1}\psi^{1})           \nonumber \\
    K=\psi_{0}\psi^{0}+\frac{k}{k+2}\psi_{1}\psi^{1}-
    \frac{\imath}{k+2}(J_{0}-J^{0}) \ee.\n

\bf  3. Physical observables in topological phase of N=2 supersymmetric WZNW
models \rm   \large
\bf  and semi-infinite cohomology of current algebras. \rm   \large

Topological phase of N=2 supersymmetric WZNW model
has the stress tensor $\tilde{T}=T+\frac{\imath}{2}\partial K$,
$\tilde{\bar{T}}=\bar{T}\pm\frac{\imath}{2}\bar{\partial}\bar{K}$.
In general there are two inequivalent models,
depending on the choice of sign in
the last formula. In this paper we will consider only
the model with the plus sign.
The model with opposite sign is belived to be
equivalent by a duality that is related
to the "mirror manifold" phenomenon [17,18].
After above-described the stress
tensor modification the oporator
\bb   Q=G^{-}(0)+\bar{G}^{-}(0)           \ee
,where $G^{-}(0)$ and $\bar{G}^{-}(0)$
are zero-modes of the holomorphyc and anti\-
holomorphyc spin-3/2 currents, can be considered
as the BRST operator. The physical
states are defined by the condition that
tey are Q-closed and equivalent up to Q-exact
states:
\bb   Qx=0
      \sim x+Qy                                        \ee
For brevity we will consider only holomorphyc
part of topological phase (the
total physical space is the tensor product holomorphyc
and anti-holomorphyc parts).
That is rather then (12) we consider
\bb   G^{-}(0)x=0
      \sim x+G^{-}(0)y                                 \ee
,where $x$, $y$ are any states from the
holomorphyc sector of N=2 supersymmetric model.
For simplicity we consider conditions (13) in the case (3),(8).\n
Let $L_{(l,k)}$ be ir\-re\-du\-ci\-ble
$\hat{Lsl(2)}$
module with integral highest weight
$(l,k)$ and highest vector $v_{(l,k)}$.
Let $F_{p,k+2}$ be ir\-re\-ducible rep\-re\-sen\-tation
of the Heisenberg algebra with generators $P(n)$, $n\in Z$
and vacuum vector $u_{p,k+2}$
annihilated by $P(n)$, $n\geq 1$ and
\bb P(0)u_{(p,k+2)}=\frac{p}{\sqrt{2(k+2)}}u_{(p,k+2)}  \ee
Let $Cl$ be ir\-re\-du\-ci\-ble rep\-re\-sen\-ta\-tion
of Clif\-ford algebra with ge\-ne\-ra\-tors
$\psi^{a}(n)$, $\psi_{a}(n)$, $a=0,1$, $n\in Z$ and vacuum
vector $\omega$ an\-ni\-hi\-lated
by $\psi^{a}$, $n\geq 1$ and $\psi_{a}(n)$, $n\geq 0$.
Consider the space
\bb A_{(l,p,k)}=L_{(l,k)}\otimes F_{(p,k+2)}\otimes Cl    \ee
In view of commutatin relations of N=2 Virasoro superalgebra
and Sugavara construction (10)
the zero mode $G^{-}(0)$ endow $A_{(l,p,k)}$ with the structure
of semi-infinite cochain complex
\bb   A_{(l,p,k)}=\oplus _{n\in Z}A^{\frac{\infty}{2}+n}             \ee
with the differential $d=G^{-}(0)$. Let $H^{\frac{\infty}{2}+\ast}
(d,L_{(l,k)}\otimes F_{(p,k+2)})$ be
semi-infinite cohomology of complex $A_{(l,pk)}$.
It is clear that all physical states coming from
$A_{(l,pk)}$ compose its cohomology.\n
The above-described considerations of the
simplest case (3),(8) can be easly
extended to more general Manin triples
(3), (6), (7).\n

\bf  4. Computation of the cohomology groups. \rm   \large

For spin-3/2 current $G^{-}(z)$ there is a decomposition
\bb   G^{-}=G^{-}_{0}+G^{-}_{1}   \nonumber \\
      G^{-}_{0}=-\sqrt{\frac{2}{k+2}}\psi^{0}(J_{0}
      +\imath\psi^{1}\psi_{1}) \nonumber \\
      G^{-}_{0}=-\sqrt{\frac{2}{k+2}}\psi^{1}J_{1} \ee
and relations
\bb   G^{-}_{0}(z)G^{-}_{0}(w)=o(z-w) \nonumber \\
      G^{-}_{1}(z)G^{-}_{1}(w)=o(z-w) \nonumber \\
      G^{-}_{0}(z)G^{-}_{1}(w)=o(z-w) \ee
Let's introduce bigrading on $A_{(l,p,k)}$ putting
\bb   deg(J^{a}(n))=deg(J_{a}(n))=(0,0) \nonumber \\
      deg(\psi^{0}(n))=-deg(\psi_{0}(n))=(-1,0) \nonumber \\
      deg(\psi^{1}(n))=-deg(\psi_{1}(n))=(0,-1) \ee
In view of (17),(18),(19) zero modes $G^{-}_{0}$, $G^{-}_{1}$
endowes $A_{(l,p,k)}$
with the structure of double complex
\bb   A^{\ast,\ast}_{(l,p,k)}=\oplus_{n,m\in Z} A^{n,m}_{(l,p,k)}   \ee
with differentials $d_{0}=G^{-}_{0}(0)$, $d_{1}=G^{-}_{1}(0)$,
$deg(d_{0})=(-1,0)$,
$deg(d_{1})=(0,-1)$. There is the spectral
sequence
\bb \{E^{n,m}_{r},\delta_{r} ,r=1,2,...\}                           \ee
associated with double complex (20), such that
\bb E^{n,m}_{1}=
    H^{\frac{\infty}{2}+m}(d_{1},A^{n,\ast}_{(l,p,k)}) \nonumber \\
    E^{n,m}_{2}=H^{\frac{\infty}{2}+n}(d_{0},E^{\ast,m}_{1}) \ee
and the limit term $E_{\infty}$ of this spectral sequence gives
the cohomology
of the complex(16) with respect to $d$.\n
In our case as we will show the spectctral sequence (21)
degenerates in the
first term, that is the differential $\delta_{r}$
in $E^{n,m}_{r}$ has to be
identicaly zero for $r>1$ and $E^{n,m}_{2}=E^{n,m}_{\infty}$.
Therefore inaccording to our rules, first of all
we have to compute $E^{n,m}_{1}$.\n
Let $Cl_{0}$, $Cl_{1}$ be irreducible representations
of Clifford algebras with
generators $\psi^{0}(n)$, $\psi_{0}(n)$ and with
generators $\psi^{1}(n)$, $\psi_{1}(n)$,
correspondingly. It is obvious that
\bb A^{\ast,\ast}_{(l,p,k)}=A^{\ast}_{0}\otimes A^{\ast}_{1} \nonumber \\
    A^{\ast}_{0}=F_{(p,k+2)}\otimes Cl_{0} \nonumber \\
    A^{\ast}_{1}=L_{(l,k)}\otimes Cl_{1} \ee
Becouse $A^{\ast}_{1}$ is the complex with differential $d_{1}$
wich act by
identity on $A^{\ast}_{0}$ we have:
\bb E^{\ast,m}_{1}=A^{\ast}_{0}\otimes H^{\frac{\infty}{2}+m}(d_{1}
,A^{\ast}_{1}) \ee
Let's denote
\bb I_{0}=J_{0}+\imath\psi^{1}\psi_{1} \nonumber \\
    I^{0}=J^{0}+\imath\psi_{1}\psi^{1} \nonumber \\
    I_{-}=I_{0}-I^{0} \nonumber \\
    I_{+}=I_{0}+I^{0}     \ee
One has:
\bb [d_{1},I_{0}(n)]=[d_{1},I^{0}(n)]=0                              \ee
{}From (23),(25),(26) follows that $E_{1}^{n,m}$ is a
representation of loop algebra
$\hat{Lh_{-}}\oplus \hat{Lh_{+}}$ generated by currents
$I_{-}(z)$, $I_{+}(z)$.
The cohomology $H^{\frac{\infty}{2}+m}(d_{1},A^{\ast}_{1})$
is given by the semi-
infinite Borel-Weil-Bott theorem:
\bb H^{\frac{\infty}{2}+m}(d_{1},A^{\ast}_{1})=
    F_{(w_{m}\ast l,k+2)}                           \ee
,where
\bb w_{m}= \left\{\begin{array}{ll}
                    \underbrace{w_{1}w_{0} \cdots w_{(0,or 1)}}_{m} &
\mbox{$m>0$} \\
                    \underbrace{w_{0}w_{1} \cdots w_{(0,or 1)}}_{-m} &
\mbox{$m<0$}
                  \end{array}
           \right. \ee
and $w_{0}$, $w_{1}$ are simple reflection in affine Weil group of
$\hat{Lsl(2,C)}$.
This theorem was prooved in [18].\n
Let $\Gamma_{0}$ be Heisenberg superalgebra with generators
$I^{0}(n)$, $I_{0}(n)$,
$\psi^{0}(n)$, $\psi_{0}(n)$. As a consequence of
the semi-infinite Borel-Weil-Bott
theorem and (24) one gets:
\bb E^{\ast,m}_{1}=
    F_{(w_{m}\ast l,k+2)}\otimes F_{(p,k+2)}\otimes Cl_{0} \ee
and $E^{\ast,m}_{1}$ is the representation of
$\Gamma_{0}$ with the vacuum
vector $v_{(l_{m},p)}$ annihilated by
$\psi^{0}(n)$, $I^{0}(n)$,$I_{0}(n)$,
with $n>0$ and $\psi_{0}(n)$ with $n\geq 0$,
and zero modes act as following:
\bb I_{-}(0)v_{(l_{m},p)}=l_{m}v_{(l_{m},p)} \nonumber\\
    I_{+}(0)v_{(l_{m},p)}=pv_{(l_{m},p)}   \ee
,where $l_{m}=w_{m}\ast l$.
Now we have to compute the second term $E^{n,m}_{2}$
of spectral sequence (21).
There are relations:
\bb [d_{0},\psi^{0}(n)]_{+}=0 &
    [d_{0},I^{0}(n)]=-\sqrt{2(k+2)}n\psi^{0}(n) \ee
\bb [d_{0},\psi^{0}(n)]_{+}=
    -\imath\sqrt{\frac{2}{(k+2)}}I_{0}(n) &
    [d_{0},I_{0}(n)]=0 \ee
The relations in (32) means that
$\psi_{0}(0)$ is the contracting
homotopy operator for $d_{0}$. Using this fact it is
not difficult to show that:
\bb E^{n,m}_{2}=
    E^{n,m}_{2,rel}\oplus \psi^{0}(0)\otimes E^{n,m}_{2,rel} \ee
,where $E^{n,m}_{2,rel}$ is $d_{0}$-cohomology of the
relative complex $E^{\ast,m}_{1,rel}$:
\bb E^{\ast,m}_{1,rel}=
    \{c\in E^{\ast,m}_{1}|\psi_{0}(0)c=I_{0}(0)c=0 \} \ee
The cohomology $E^{n,m}_{2,rel}$ is very simple
becuose they are cohomology of Heisenberg algebra
generated by $I_{0}(z)$. We have:
\bb E^{n,m}_{2,rel}= \left\{\begin{array}{ll}
                              0 & \mbox{$n\neq 0$} \\
                              C & \mbox{and generated by $v_{(l_{m},p)}$ if}
                            \end{array}
                     \right. \ee
\bb I_{0}(0)v_{(l_{m},p)}=0       \ee
{}From this result follows that the spectral
sequence (21) degenerates at the
first term $E^{n,m}_{2}=E^{n,m}_{\infty}$
and one gets the following:
\bb H^{\frac{\infty}{2}+\ast}(d,L_{(l,k)}\otimes F_{(p,k+2)})= \nonumber \\
    H^{\frac{\infty}{2}+\ast}_{rel}(d,L_{(l,k)}\otimes F_{(p,k+2)})
    \oplus    \nonumber \\
      \psi^{0}(0)\otimes H^{\frac{\infty}{2}+\ast}_{rel}(d,L_{(l,k)}\otimes
F_{(p,k+2)})
\ee
\bb H^{\frac{\infty}{2}+n}_{rel}(d,L_{(l,k)}\otimes F_{(p,k+2)})=
\delta_{n,m}\delta_{(l_{m}+p+1)}C                                    \ee
,where $H^{\frac{\infty}{2}+\ast}_{rel}(d,L_{(l,k)}\otimes F_{(p,k+2)})$
denote $d$-cohomology of the relative complex
\bb A^{\ast}_{rel}=\{c\in A_{(l,p,k)}|\psi_{0}(0)c=I_{0}(0)c=0\}  \ee
and delta-function $\delta_{(l_{m}+p+1)}$ code the condition (36).\n

The generalization of the last formulas for Manin triples
(3), (6), (7) is
straightforward becouse
in [19] semi-infinite Borel-Weil-Bott theorem
proved for general case.\n

\bf  5. Representatives of physical states  \rm   \large
\bf and ground states in Ramound sector. \rm   \large \n
Now we give in an explicit form a representatives of
$H^{\frac{\infty}{2}+\ast}(d,L_{(l,k)}\otimes F_{(p,k+2)})$.\n
Let us suppose that $H^{\frac{\infty}{2}+m}_{rel}$ is nonzero.
As is clear from calculations (29) ,(35) of spectral sequence
the representative $H^{\frac{\infty}{2}+m}_{rel}$ is vacuum vector
$v_{(l_{m},p)}$ of Heisenberg superalgebra $\Gamma_{0}$. But in view of
semi-infinite Borel-Weil-Bott theorem this representative
is in one to one correspondence with vacuum vector out of
$H^{\frac{\infty}{2}+m}(d_{1},A^{\ast}_{1})$.
Therefore in order for the representative
of $H^{\frac{\infty}{2}+m}_{rel}$
to write out suffice it to write out the vacuum vector
$v_{l_{m}}\in H^{\frac{\infty}{2}+m}(d_{1},A^{\ast}_{1})$
and tensor multiply its by fermion vacuum
$\omega_{0}$ of $Cl_{0}$ and by vacuum vector $u_{(p,k+2)}$
of $F_{(p,k+2)}$ ,where momentum $p$ is defined from (36).\n
For any ghost number $m$ the vacuum vector of
$H^{\frac{\infty}{2}+m}(d_{1},
A^{\ast}_{1})$ can be obtained as follows. Let $v_{(l,k)}$
be highest vector of $L_{(l,k)}$ and $\omega_{1}$
be vacuum vector of $Cl_{1}$.
The direct calculation shown that
$v_{(l,k)}\otimes \omega_{1}$ is vacuum
vector of $H^{\frac{\infty}{2}+0}(d_{1},A^{\ast}_{1})$.
It is possible to prove
that vacuum vector with nonzero ghost number $m$
is given by the action of the
element $w_{m}$ out of affine Weil group
of $\hat{Lsl(2)}$ on vacuum
$v_{(l,k)}\otimes \omega_{1}$ ($w_{m}$ is determined in (28)).\n
In ac\-cord\-ing to this pro\-ce\-dure we can
to write out the rep\-re\-sen\-ta\-tive
$H^{\frac{\infty}{2}+m}_{rel}$ for any
ghost namber $m$. Recall that $\omega$
is the vacuum out of $Cl$
and let $n$ be a nonnegative integer number.Then
\bb X_{-2n}=\psi^{1}(-2n+1)\psi^{1}(-2n+2)\cdots
    \psi^{1}(0)\omega\otimes \nonumber \\
    (w_{1}w_{0})^{n}v_{(l,k)}\otimes u_{(-1-l_{-2n},k+2)} \nonumber \\
    X_{-2n-1}=\psi^{1}(-2n)\psi^{1}(-2n+1)\cdots
    \psi^{1}(0)\omega\otimes \nonumber \\
    w_{1}(w_{1}w_{0})^{n}v_{(l,k)}\otimes u_{(-1-l_{-2n-1},k+2)} \nonumber \\
    X_{2n+1}=\psi_{1}(-2n-1)\psi_{1}(-2n)\cdots
    \psi_{1}(-1)\omega\otimes \nonumber \\
    (w_{0}w_{1})^{n}w_{0}v_{(l,k)}\otimes u_{(-1-l_{2n+1},k+2)} \nonumber \\
    X_{2n}=\psi_{1}(-2n)\psi_{1}(-2n+1)\cdots
    \psi_{1}(-1)\omega\otimes \nonumber \\
    (w_{0}w_{1})^{n}v_{(l,k)}\otimes u_{(-1-l_{2n},k+2)}
\ee
be representatives with ghost nambers $-2n$ ,$-2n-1$, $2n+1$, $2n$
correspondingly.Using this explicit formulas and properties of the
affine Weil group representation
on the $L_{(l,k)}$ [20] it is possible
to show that a representatives (40) are ground states
in Ramound sector of the
N=2 Virasoro superalgebra,
that is they annihilated by positive modes of
currents $K(z)$, $G^{\pm}(z)$, $T(z)$
and the action of zero modes is given by
\bb G^{\pm}(0)X_{m}=0              \ee
\bb L(0)X_{m}=\frac{c}{24}X_{m}    \ee
\bb K(0)X_{m}=(\frac{l}{k+2}-\frac{c}{6})X_{m}
    & \mbox{if $m$ is even} \nonumber\\
    K(0)X_{m}=(\frac{k-l}{k+2}-\frac{c}{6})X_{m}
    & \mbox{if $m$ is odd} \ee
In order for a representatives of
$H^{\frac{\infty}{2}+\ast}(d,L_{(l,k)}\otimes F_{(p,k+2)})$
to write out and compute its U(1)-charges suffice
it to use a representatives of
$H^{\frac{\infty}{2}+m}_{rel}$ and formula (37).
It's easy to see that they are
also ground states in Ramound sector.\n
The representative constructing procedure
may be easily extended for Manin
triples (3), (6), (7).\n

\bf  6. The ring structure of  physical states. \rm   \large  \n
In the last section a correspondence
between cohomology classes and ground states
in Ramound sector was established. The spectral
flow of N=2 Virasoro superalgebra [21]
connects the Ramound (R) and Neveu-
Schwarz (NS) sectors. Under this flow the
ground states of R-sector flow to the
chiral primary fields of ns-sector. U(1)-
charges of chiral primary fields differs from
U(1)-charges of ground states by $-\frac{c}{6}$
and from (43) we conclude that U(1)-charges
of chiral primary correspondings to the
representatives of
$H^{\frac{infty}{2}+\ast}_{rel}$ belongs to
the segment of unitarity bound of N=2
Virasoro superalgebra [22-23]:
\bb g_{\frac{1}{2}}=0 & 0\leq q \leq \frac{c}{3}-1 \ee
,where $q$ is U(1)-charge. U(1)-charges of chiral
primary fields correspondings to the representatives
of $\psi^{0}\otimes H^{\frac{\infty}{2}+\ast}_{rel}$
belongs to the segment of unitarity bound:
\bb g_{\frac{1}{2}}=0 & 1\leq q \leq\frac{c}{3} \nonumber \\
    g_{\frac{3}{2}}<0 & f_{1,2}\geq 0      \ee .\n
The operator product of chiral primary fields
modulo setting to zero the descandants of chiral
primary fields defines the ring stucture of
of chiral primary fields and thus- the ring
structure on the cohomology classes [21].\n
Let's denote the conformal fields correspondings
to the vectors $v_{(l,k)}$, $u_{(p,k+2)}$ by
the same letters: $v_{(l,k)}(z)$, $u_{(p,k+2)}(z)$.
Then the chiral primary fields correspondings
to the representatives of the relative
cohomology classes are given by:
\bb X_{-2n,l}=\partial^{2n-1}\psi^{1}\partial^{2n-2}\psi^{1}\cdots
    \psi^{1}\otimes \nonumber \\
    (J^{1}(-2n+1))^{k-l}(J^{1}(-2n+2))^{l}\cdots \nonumber \\
    (J^{1}(0))^{k-l}v_{(l,k)}\otimes u_{(-l+2n(k+2),k+2)} \nonumber \\
    X_{-2n-1,l}=\partial^{2n}\psi^{1}\partial^{2n-1}\psi^{1}\cdots
    \psi^{1}\otimes \nonumber \\
    (J^{1}(-2n))^{l}(J^{1}(-2n+1))^{k-l}\cdots \nonumber \\
    (J^{1}(0))^{l}v_{(l,k)}\otimes u_{(l+2+2n(k+2),k+2)} \nonumber \\
    X_{2n+1,l}=\partial^{2n}\psi_{1}\partial^{2n-1}\psi_{1}\cdots
    \psi_{1}\otimes \nonumber \\
    (J_{1}(-2n-1))^{k-l}(J_{1}(-2n))^{l}\cdots \nonumber \\
    (J_{1}(-1))^{k-l}v_{(l,k)}\otimes u_{(l+2-2(n+1)(k+2),k+2)} \nonumber \\
    X_{2n,l}=\partial^{2n-1}\psi_{1}\partial^{2n-2}\psi_{1}\cdots
    \psi_{1}\otimes \nonumber \\
    (J_{1}(-2n))^{l}(J_{1}(-2n+1))^{k-l}\cdots \nonumber \\
    (J_{1}(-1))^{k-l}v_{(l,k)}\otimes u_{(-l-2n(k+2),k+2)} \nonumber \\
\ee
The rest chiral primary are given by maltiplication
of the above on the chiral primary field $\psi^{0}$.
The ring structure of the chiral primary fields is
restricted by Ward identities and may be described
as follows. Let's denote $X_{-1,k}=\theta_{1}$,
$\psi^{0}=\theta_{2}$, $X_{0,1}=y$, $X_{2,0}=x$,
$X_{-2,0}=x^{-1}$, $X_{0,0}=1$. Then the
chiral ring is the associative supercommutative
ring mith unit generated by $\theta_{1}$,
$\theta_{2}$, $x$, $x^{-1}$, $y$ and relations:
\bb [\theta_{1},\theta_{2}]_{+}=\theta_{1}^{2}=
    \theta_{2}^{2}=0 \nonumber \\
    xx^{-1}=x^{-1}x=1 \nonumber \\
    y^{k+1}=1
\ee. \n

This work was supplied in part, by Soros Foundation Grant, Awarded by
the American Physical Society. I am deeply grateful to B. Feigin for
discussions.

\vspace{5.00mm}
 \underline{\bf References} \rm \n
 [1] Gepner,Nucl. Phys. B296, 757 (1988);
     Phys. Lett. 199B, 380 (1987). \n
 [2] T.Eguchi, H.Ooguri, A.Taormina and S-K.Yang,
     Nucl. Phys. B315, 193 (1989). \n
 [3] A.Taormina, CERN-preprint (May 1989). \n
 [4] Y.Kazama, H.Suzuki, Mod. Phys. Lett. A4, 235 (1989);
     Phys. Lett. 216B, 112 (1989);
     Nucl. Phys. B321, 232 (1989). \n
 [5] K.Li, Calteh-preprint CALT-68-1662.\n
 [6] T.Eguchi, S-K.Yang, Tokyo-preprint UT-564. \n
 [7] E.Witten, Nucl. Phys. B340, 281 (1990). \n
 [8] E.Martinec, Phys. Lett. B217, 431 (1987); \n
     C.Vafa, N.Warner, Phys. Lett. B218, 51 (1989). \n
 [9] P.Spindel, A.Sevrin, W.Troost, A.Van Proeyen,
     Nucl. Phys. B308, 662 (1988);
     B311, 465 (1988/89). \n
 [10] C.M.Hull,B.Spence, Queen Marry Callege-preprint \n
      QMC/PH/89/24. \n
 [11] S.Parkhomenko, Zh. Eksp. Teor. Fiz. 102, 3-7 (July 1992). \n
 [12] M.Gunaydin, J.L.Petersen, A.Taormina, A.Van Proeyen,
      Nucl. Phys. B322, 402 (1989). \n
 [13] J.L.Petersen, A.Taormina, CERN-TH.5446/89. \n
 [14] J.L.Petersen, A.Taormina, EFI-90-61, August 1990. \n
 [15] J.L.Petersen, A.Taormina, CERN-TH.5503/89. \n
 [16] H.Ooguri, J.L.Petersen, A.Taormina,
      Nucl. Phys. B368, 611 (1992). \n
 [17] P.Candelas, M.Lyker, R.Schimmrigk,
      Nucl. Phys. B341, 383 (1990). \n
 [18] C.Vafa, HUTP-91/A 059 \n
 [19] B.Feigin, E.Frenkel, Commun.Math. Phys. v.128, 161 (1990). \n
 [20] V.G.Kac, Infinite Di\-men\-si\-onal
      Lie Algebras,\n
      Birkhaser,
      Boston-Basel-Stuttgart,1983. \n
 [21] W.Lerhe, C.Vafa, N.Warner, HUTP-88/A065.\n
 [22] W.Boucher, D.Friedan, A.Kent, Phys. Lett. B172, 316 (1986).\n
 [23] V.K.Dobrev, Phys. Lett. B186 43 (1987).\n


\end{document}